%% file: webmodels-xp-report.tex
\documentclass{article}

\usepackage{style/arxiv}

\usepackage[utf8]{inputenc} 
\usepackage[T1]{fontenc}    
\usepackage[hidelinks]{hyperref}       
\usepackage{url}            

\usepackage[acronym,toc]{glossaries}
\loadglsentries[\acronymtype]{acronyms}
\makeglossaries

\usepackage{booktabs}       
\usepackage{amsfonts}       
\usepackage{nicefrac}       
\usepackage{microtype}      
\usepackage{cleveref}       
\usepackage{lipsum}         
\usepackage{graphicx}
\usepackage{natbib}
\usepackage{doi}

\title{Implementing a Model-based Engineering Tool as Web Application}

\date{}

\author{Florian Hölzl \\
	fortiss GmbH\\
	Research Institute of the Free State of Bavaria\\
	Guerickestraße 25 \\
	80805 Munich \\
	Germany\\
	\texttt{hoelzl@fortiss.org}\\
	\And
	Simon Barner\thanks{corresponding author}\\
	fortiss GmbH\\
	Research Institute of the Free State of Bavaria\\
	Guerickestraße 25 \\
	80805 Munich \\
	Germany\\
	\texttt{barner@fortiss.org}
}

\renewcommand{\headeright}{}
\renewcommand{\undertitle}{Experience Report}

\hypersetup{
pdftitle={Implementing a Model-based Engineering Tool as Web Application},
pdfsubject={Experience Report},
pdfauthor={Florian Hölzl, Simon Barner},
pdfkeywords={Model-based Engineering, Tooling, Web Technology},
}

\begin{document}
\maketitle

\begin{abstract}
This paper reports on a study of transferring a desktop-based model-based engineering tool to a web application. The study has been conducted in the WEBMODEL project where the well-established technology stack around the Eclipse platform and the Eclipse Modeling Framework was lifted into a cloud-based environment. As results, a modeling language independent tooling kernel for web-based modeling tools and a minimal prototypical web-based implementation of the AutoFOCUS~3 model-based engineering tool are presented. Furthermore, the report documents experiences and implementation advises gained during the implementation.
\end{abstract}


\section{Introduction}
This paper reports on the work conducted in the WEBMODEL project (see Section~\ref{sec:acknowledgment}) that aimed at showing the feasibility of bringing modeling tools, which are usually implemented as traditional desktop applications, to web applications. The project was carried out by members of EclipseSource Munich\footnote{\url{https://eclipsesource.com/about/company/eclipsesource-munich/}} and the Model-based Systems Engineering field of competence of the fortiss research institute\footnote{\url{https://www.fortiss.org/en/research/fields-of-research/detail/model-based-systems-engineering/}}. In the project, the partner EclipseSource took the role as a provider of the technologies and frameworks to enable web-based modeling tools, while fortiss was responsible for in-depth testing, bug-reporting, evaluation and improving the design and implementation of the resulting framework. The evaluation has been conducted in a case study where an \gls{MVP} of the simplistic model-based engineering tool has been implemented based on the WEBMODEL framework.

\section{Approach}
In the WEBMODEL project, the well-established technology stack around the Eclipse platform and the \gls{EMF} was lifted into a cloud-based environment with the idea of reusing existing components of modeling tools, which are implemented using the former technology and making them available to their users via a web-browser interface. The transition to a web-based tool environment was realized by couple of newly implemented base technologies: some backend servers implemented in Java as Eclipse products and some frontend frameworks implemented in JavaScript / TypeScript as NodeJS\footnote{\url{https://nodejs.org}} modules.

This document provides the details and collects the experience gained during the prototypical implementation of a simplistic version of the modeling features of the  \gls{AF3} model-based systems tool using the web- and cloud-based technologies developed by the WEBMODEL research project.

\subsection{AutoFOCUS~3}
AutoFOCUS~3 is an open-source model-based systems engineering tool and research platform for safety-critical embedded systems. It is a desktop tool that builds on the Eclipse platform and the \gls{EMF}. \gls{AF3} has been implemented following the SPES modeling approach and engineering methodology (see~\cite{Pohl2012}). \gls{AF3} enables graphical modeling of designs and implementations of safety-critical systems in terms of the SPES modeling viewpoints (see ~\cite{Aravantinos2015} and~\cite{Barner2018}). Besides a model of the vertical and horizontal structure, all viewpoints provide a rich set of annotations that are mainly used to describe non-functional properties of the respective model element.

The tool provides advanced features that enable to explore design and implementation alternatives (e.g., safety patterns, task allocation, partition/compartment architectures) based on state-of-the art formal methods and solvers (e.g., Z3 solver), and to validate early designs by means of functional simulation and co-simulation (via FMI). The automatic artefact synthesis capabilities of the tool enable rapid development (e.g., code and configuration generation, scheduling synthesis). Modular assurance cases encoded in the goal-structuring notation (GSN) enable users to document the dependability of a system by bringing in information about its environment and the development context. They provide a structured argument that establishes a systematic relationship between a dependability goal and its evidence (e.g., validation artefact).

\subsection{Case Study}
In order to evaluate the applicability of the WEBMODEL framework, a reduced version of the Eclipse-based \gls{AF3} desktop tool has been implemented (the so-called \gls{MVP}). It is intended to show-case the web-based modeling technology created in the WEBMODEL project. The advanced features of the \gls{AF3} summarized above are beyond the scope of the \gls{MVP}.
The \gls{MVP} uses a simplified metamodel compared to its desktop counterpart, which is still representative with respect to the art of metamodeling. There are three major parts represented in this metamodel: first, a simple language for modeling requirements mainly using form-based editing capabilities; second, a graphical language for modeling component architectures similar to the \gls{AF3} editors; thirdly, an allocation model between requirements and components, which maps requirements to these components (implementation mapping).

The different parts of the \gls{MVP} are used to demonstrate the different parts of the WEBMODEL technology stack (described in detail in the next section). The requirements model primarily shows the hierarchic modeling and form-based modeling capabilities. Hierarchic modeling is also shown by the component architecture modeling functionality. Furthermore, each level of the hierarchy can be depicted and edited by using a graphical editor, which is implemented using the \gls{GLSP} framework.
In the frame of the case study, only a small subset of the reduced metamodel is actively used. No full-fledged \gls{CRUD} support is implemented for each metamodel entity. However, all necessary pieces of technology have been included in at least some part of the \gls{MVP}, which makes the support of the complete \gls{CRUD} set merely a matter of time and resources. The \gls{CTK4} infrastructure (see Section~\ref{sec:ctk4}) provides abstract base classes that foster and ease the implementation of a fully functional modeling tool based on the WEBMODEL technologies.

\section{Web-based Modeling Tool Architecture} 
In the following, we describe the system architecture of the minimal viable product used to create a browser-based modeling tool based on the WEBMODEL technology.

\begin{figure}[!t]
	\centering
	\includegraphics[width=\textwidth]{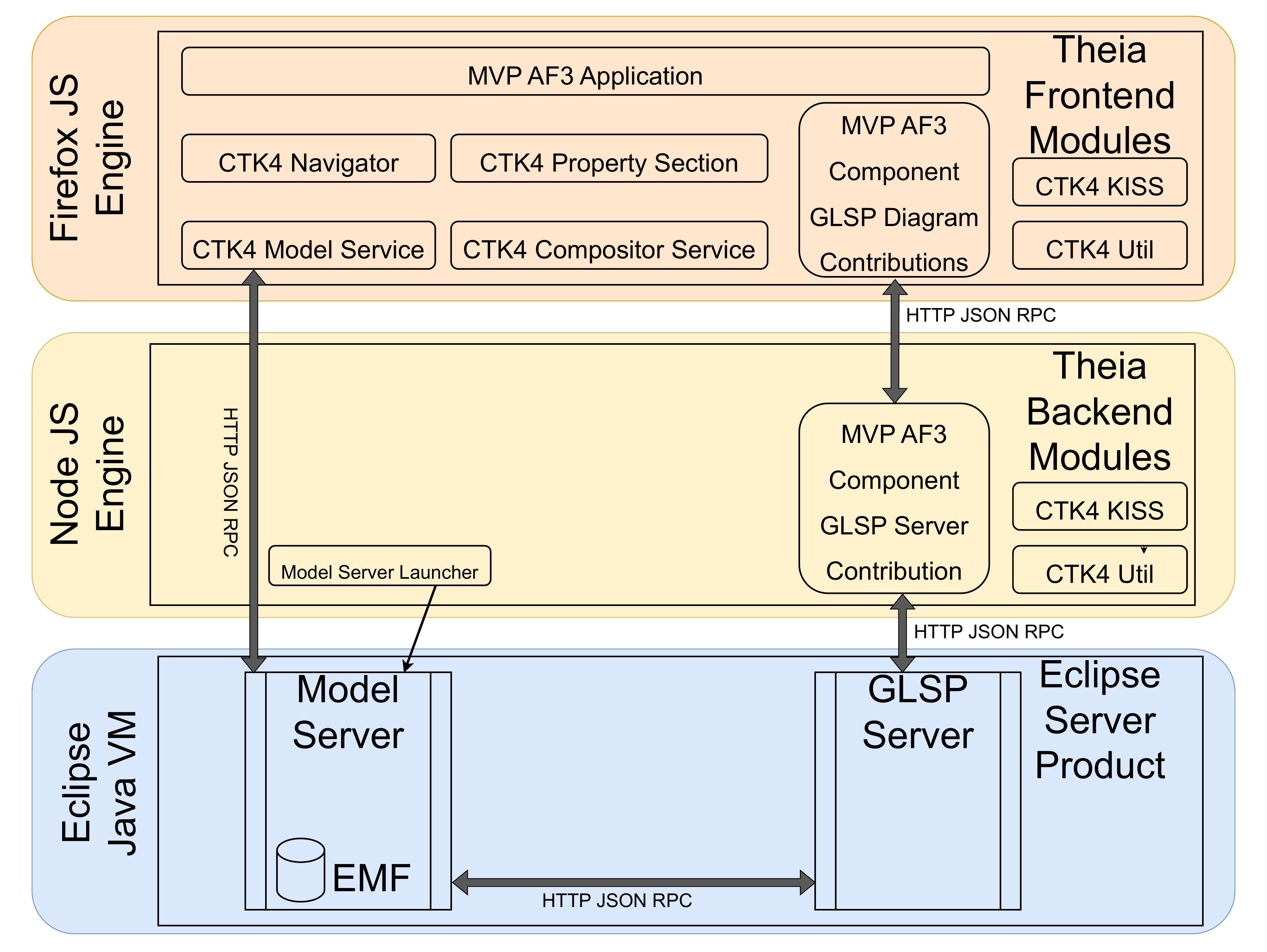}
	\caption{Overview of WEBMODEL System Architecture.}
	\label{fig:architecture-overview}
\end{figure}

Figure~\ref{fig:architecture-overview} illustrates the overall architecture that consists of three layers, namely the Eclipse Server Product, the Theia backend modules as well as the Theia frontend modules. It also depicts the main components of each  layer, as well as the communication paths and the used communication protocols. Theia\footnote{\url{https://theia-ide.org/}} is a browser application framework implementing the basic services, which are also provided by the Eclipse framework for pure desktop-based tools. It uses and extends the Webpack\footnote{\url{https://webpack.js.org/}} module packer engine, a Javascript module packer used to generate bundled and therefore more efficient JavaScript applications. Theia uses Webpack to produce the executable compressed application archives run in the NodeJS runtime.
The left-hand side of the figure shows the execution environments of the different layers, namely a Java virtual machine and the frontend and backend JavaScript engine.
Some modules are split into frontend and backend modules within the Theia layers of the architecture. The server implementations are done in Java and run as Eclipse products.

\subsection{Theia Frontend}
The Theia Frontend is a browser application that provides the main interface for the tool user for editing and analyzing the model under construction. It is built on top the Theia framework, which provides services similar to the Eclipse platform services used in traditional desktop modeling tools. It uses and extends the Webpack module packer engine.

Theia provides, for example, services to access the file system or the user’s workspace, to display messages or to handle a global selection across different views. The framework also provides the mechanisms to define editors and views, again similar to the Eclipse workbench API.

\begin{figure}[!t]
	\centering
	\includegraphics[width=\textwidth]{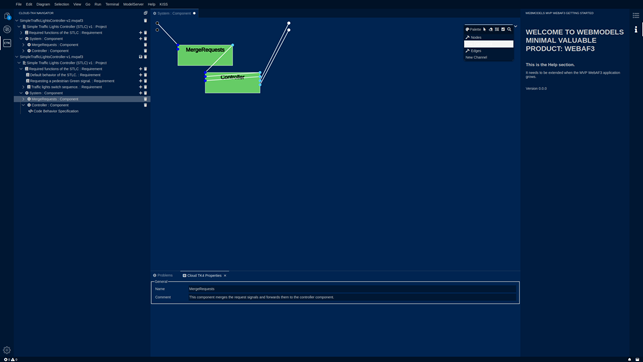}
	\caption{Graphical editors of AF3 MVP.}%
	\label{fig:theia-fullscreen}
\end{figure}

Using these mechanisms, the \gls{MVP} implemented three major parts (as depicted in Figure~\ref{fig:theia-fullscreen}): a model navigator to the left, graphical editors for depicting more complex diagrams in the center, and form-based editors for simple model properties (like element names and comments) in the lower part of the figure. The \gls{MVP} also contains an introductory welcome page, which is implemented as a view on the right of the application window.

\paragraph{Model Tree Navigator}
The model tree navigator displays the hierarchic structure of all models provided by the model server. Selecting an element in the navigator shows its properties in the model property section below the main editor area. The interaction symbols depicted to the right of each line in the model navigator allow to add and remove elements and therefore edit the tree structure. Next to the model file name the floppy disk symbol allows to issue the save command, which persists the model state on the model server.

\paragraph{Graphical Editors}
The center area of the application window is reserved for the main editors (textual, e.g., code, or graphical, e.g., diagrams). The \gls{MVP} only contains a single type of editor for modeling component architectures, which are depicted by diagrams of connected graphs. Graphical editors show the diagram and provide a tool palette with different means to add elements to the diagram or remove them from it. In the \gls{MVP} example, the palette is minimal, since only the adding of component nodes has been implemented.

\paragraph{Form-based Editing of Model Properties}
The lower part of the application window is occupied by the model properties view, which allows to edit the properties of model elements selected in the tree navigator. These properties are usually attributes like name or comment or other simple key-value data items. Each entry in the property section can display error messages when a custom validation of the entered data fails. A typical example would be the syntactic validity check of an email address.

\paragraph{Theia Frontend Modules}
The parts of the \gls{MVP} application architecture discussed so far are implemented as Theia frontend modules using the TypeScript programming language. In order to define such modules, the \texttt{package.json} file usually contains three parts: a dependency to \texttt{@theia/core}, an entry in the keyword section named \texttt{theia-extension}, and the pointer to the frontend module implementation, e.g., \texttt{kiss.frontend.module}. 

\subsection{Theia Backend Modules}
Theia backend modules are defined very similar to their frontend counter-parts. The dependency, the keyword and a pointer to the backend module implementation is all that is needed to specify a Theia backend module. The main purpose of those backend modules is to launch auxiliary processes, like the model server or the \gls{GLSP} server, if they are not yet running once the Theia NodeJS environment is started (there is an auto-detection mechanism to see if the respective server is already up).

However, Theia backend modules are rarely used in the \gls{MVP}, since the servers are launched via Eclipse run configurations directly from the development Eclipse instance. A release version would include such launchers to ease the deployment.

In order to understand Theia backend modules and their implementation, the \gls{KISS} (see Section~\ref{sec:ctk4}) is a good starting point. It shows how to implement a service in the \gls{CTK4} infrastructure that needs to be available both from the frontend part as well as the backend.

\subsection{Model Server and EMF Metamodels}
The primary entity under development in a model-based software engineering tool is the model, which is automatically transformed into code-centric implementation languages during later stages of the system development process. Therefore, in the context of a web- and model-based development tool, the model must be handled by an application server, which provides the model data over a data connection. This work is carried out by the Model Server , the main implementation result of the WEBMODEL project. This server provides a remote procedure call interface accessible via HTTP connections. In the context of the \gls{MVP} prototype the model server is running on \texttt{localhost} under port 8081, e.g., accessing \texttt{http://localhost:8081/api/v1/modeluris} returns a JSON datagram listing the URIs of all models currently available on the model server.

In the world of models, metamodeling is the de-facto method to define the syntactic structure of models used in model-based development. The \gls{EMF} is a widely used framework for defining and managing metamodels as well as dealing with persistency, migration, and versioning of models for model-based tools built on top of the Eclipse platform. The metamodels used in the web-based version of a modeling tool can be reused from an existing desktop based predecessor and are plugged into the model server with only a couple of lines of glue code.

\subsection{Model Validation Framework}
\begin{figure}[!t]
	\centering
	\includegraphics[width=\textwidth]{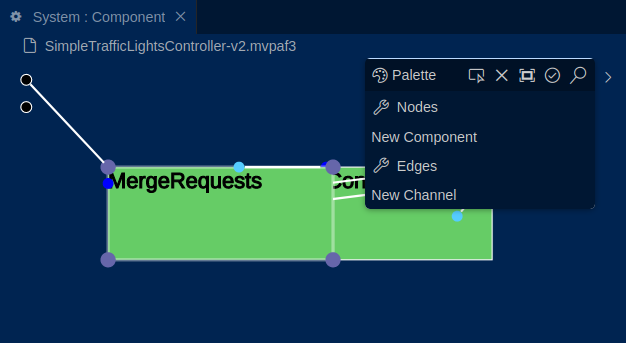}
	\caption{Triggering of complex validations from the frontend UI.}%
	\label{fig:validation-ui}
\end{figure}

Validating the model data distinguishes between simple input validations (e.g., syntactic correctness of an email address) and more complex validations (e.g., checking of a component hierarchy for cycles of weakly causal components). Simple input validation is part of property section inputs and the form-based modeling capability. 
Complex validations are triggered using the check-mark button of graphical editors as can be seen in Figure~\ref{fig:validation-ui}.

\begin{figure}[!t]
	\centering
	\includegraphics[width=0.7\textwidth]{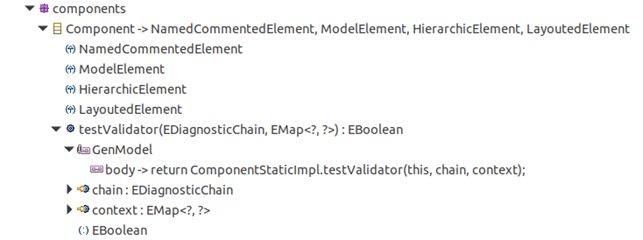}
	\caption{Implementation of validators by means of \gls{EMF} operations.}%
	\label{fig:validation-impl}
	\vspace{-0.35cm}
\end{figure}

The implementation of such validators is carried out by directly adding validation methods to the classes in the \gls{EMF} metamodel (using the \gls{EMF} Validation Framework) as depicted in Figure~\ref{fig:validation-impl}. Since this method is included in the generated code part of the metamodel plugin, we use the delegation call to \texttt{ComponentStaticImpl.testValidator(\dots)} in order to avoid including hand-written code in generated classes. This approach is well-known from the \gls{AF3} development process and is supported by the fortiss Eclipse development tools\footnote{\url{https://git.fortiss.org/af3/af3/-/wikis/Fortiss_Eclipse_Development_Tools}}.

\subsection{Graphical Language Server Platform}
The \gls{GLSP} server transforms the model and diagram data received from the model server into the format required by the Sprotty sub system, which then sends this information to the browser application for rendering the diagram as one big scalable vector graphics content. Sprotty\footnote{\url{https://projects.eclipse.org/projects/ecd.sprotty}} is a framework for displaying and editing diagrams in a browser application (e.g., a Theia view) using \gls{SVG} rendering. This conversion is carried out automatically by the framework. The implementer can focus on the operations and actions that need to be realized for a given diagram (type). In the following paragraphs, we will elaborate this procedure and its requirements.

The configuration of an instance of the framework, in other words a single type of diagram, is done by defining a \gls{GLSP} module class instance  (usually derived from \texttt{DefaultGLSPModule}). Here, next to some auxiliary definitions, four main extensions in the form of implementation classes are necessary:

\begin{itemize}
	\item A \texttt{DiagramConfiguration}  instance implements a mapping between node and edge IDs to elements of \gls{EMF} package literals. These identifiers are used to tag the elements in the final diagram, which then allows to apply CSS styling based on these IDs.
	\item A \texttt{GLSPServer}  instance implements the basic setup of the \gls{GLSP} server instance. In case of the \gls{MVP}, it also defines a communication interface to the model server.
	\item A \texttt{ModelSourceLoader}  instance receives the model state from the model server and creates a root instance of the diagram model state. It is the first part of a two-stage conversion process for the \gls{EMF} model to the graphical diagram model and is only invoked once when the diagram is opened in the Theia application.
	\item A \texttt{GModelFactory}  instance enriches the diagram model state by interpreting the result of the source loader and converting it into the graph model. This is the second part of the two-stage conversion procedure. The model factory is called every time an update to the diagram is required, e.g., adding a new node to the diagram.
\end{itemize}

\section{Cloud Tooling Kernel 4 (CTK4)}
\label{sec:ctk4}
The \gls{CTK4} transfers the concepts of the tooling kernel~\footnote{\url{https://git.fortiss.org/af3/kernel}} that has successfully been adopted by the \gls{AF3} desktop tool (see~\cite{Aravantinos2015}) to the web- and cloud-based technology used by the WEBMODEL framework.

\subsection{Metamodel}
\gls{CTK4} provides an abstract metamodel based on the \gls{EMF}, which consists of three sub-models:

\begin{itemize}
	\item a generic metamodel for hierarchic modeling languages
	\item a metamodel for layout information that can be used to model graphical information alongside the model elements of the former model, and
	\item a metamodel for diagrams, which extends the metamodel provided by the \gls{GLSP} framework.
\end{itemize}

\subsection{Model Server and GLSP Server}
There is no abstract support layer for the model server and the \gls{GLSP} server yet. The abstraction of the model server framework as well as those of the \gls{GLSP} framework are used directly by the \gls{MVP} implementation. Implementing an abstract support layer for the \gls{GLSP} framework belongs to list of possible extensions. 

\subsection{Theia Services: Base Implementations and \gls{MVP} \gls{AF3} Instantiation}

The implementation packages of the Theia modules are divided into base implementations of the \gls{CTK4} infrastructure and its concrete instantiation for the \gls{MVP} \gls{AF3} prototype. Since a lot of these services are also part of the \gls{AF3} desktop tooling kernel, a large part of the representative transfer of a desktop modeling tool to a web-based tool that was investigated in the presented case study was concerned with 1) the definition the respective generic web-based services in ~\gls{CTK4}, and 2) mapping these generic services to tool-specific implementation. In the following, each of these packages is briefly described:
	
\begin{itemize}
	\item CTK Util: contains common utility functions.
	\item CTK Introspection System Service: contains the base implementation of the debugging service, where other parts of the architecture can publish data and information during runtime. \gls{KISS} is intended to provide high-level debugging information (including current state information of other \gls{CTK4} services) in contrast to the low-level information available from the command console.
	\item CTK Metamodel: contains a basic Typescript implementation of the \gls{EMF} metamodel. For example, \texttt{EObject}, \texttt{EString} and others are defined here and used by other parts of the \gls{CTK4} infrastructure.
	\item CTK Model Service: contains an extendable service that is responsible for converting model data received from the model server into the object structure defined by the CTK metamodel implementation.
	\item CTK Model Element Handler Service: contains the implementation of the extendable service for describing typical visualizations of model elements (name, description, image).
	\item CTK Composition Service: contains the service for registering compositors that describe which model elements can be added to one another (e.g. \texttt{Component} $\leadsto$ \texttt{Sub-Component}).
	\item CTK Model Navigator View: contains the implementation of the model navigator tree-view.
	\item CTK Model Properties View: contains the implementation of the form-based model properties editing capabilities.
	\item CTK Model Properties Extension Service: contains the base implementation of the model properties framework. Providing extensions here configures what is shown in the properties view when a model element is selected.
	\item CTK Welcome Page: contains the base implementation of a welcome page.
	\item \gls{MVP} \gls{AF3} Metamodel: instantiation of the \gls{CTK4} base metamodel for the \gls{MVP} prototype.
	\item \gls{MVP} \gls{AF3} Welcome Page: instantiation of the welcome page.
	\item \gls{MVP} \gls{AF3} Theia Model Server: contains the launcher script for starting the model server when the Theia application is launched, and the model server is not already running (e.g. by having been started from Eclipse).
\end{itemize}

Based on the \gls{CTK4} kernel services described above, the following extensions have been implemented to realize the \gls{AF3} \gls{MVP} as part of the case study conducted in this work:
	
\begin{itemize}
	\item \gls{MVP} \gls{AF3} Composition Extension: contains the concrete implementation of the model element compositors (well-known from the \gls{AF3} desktop application).
	\item \gls{MVP} gls{AF3} Model Element Handler Extension: contains the concrete implementation of the model element handlers (well-known from the \gls{AF3} desktop application).
	\item \gls{MVP} \gls{AF3} \gls{GLSP} Component Diagram: contains the implementation of the component diagram editor based on the \gls{GLSP} framework and connecting to the \gls{GLSP} server usually launched from Eclipse.
	\item \gls{MVP} \gls{AF3} Theia Application: contains the main application, which references and binds the other parts together in order to form the web application listening on port 3000.
	\item \gls{MVP} \gls{AF3} Example: contains the model data, which should be symlinked into the workspace area of the model server working directory.
\end{itemize}

\section{Discussion}
In the following, some challenges regarding different aspects encountered during the implementation of the \gls{MVP} are discussed.
No signification challenges related to the backend parts of the framework and the \gls{MVP} prototype have been encountered, since the desktop version of \gls{AF3} is also an Eclipse product.
Therefore, the following discussion of challenges, including potential remedies, focuses on the frontend part of the implementation.
Some of the challenges mentioned here must be viewed in relation to the fact that over the course of the project, the framework and its implementation has improved continuously, while at the same time more and more experience in the technologies involved in the Theia part of the framework had been acquired.

\paragraph{Inconsistency of Backend and Frontend Implementations}
\textit{Challenge:} Several times during the course of the evaluation of the framework and the implementation of the \gls{MVP} prototype, inconsistencies between the backend (Java-based implementation) and the frontend (Typescript-based implementation) led to extensive development and debugging cycles. The problem is that both are not covered by a consistency checker but are independently compiled and deployed without knowledge of each other.

\textit{Advice:} the in-code documentation of frontend and corresponding backend code should reference each other. This is in particular relevant for configuration objects and ``magic'' String identifiers that have to match in both parts.

\paragraph{Dependency Injection}
\textit{Challenge:} Throughout the framework, dependency injection is used to allow for custom implementations and overrides of the default behavior of the WEBMODEL code base. During the course of implementation of the \gls{MVP} prototype and the CTK4 services, in particular, custom implementations had to be injected both into the frontend as well as the backend frameworks. Here, another inconsistency problem between injected objects arose: It was sometimes not obvious, which set of implementations had to be injected in order to achieve a consistent and properly working implementation.

\textit{Advice:} implement internal sanity checks for the case when the framework is customized using injected objects.

\section{Conclusion}
The target of this work was to evaluate the transfer of the existing desktop tool \gls{AF3} into a web-based modeling tool. In this section, the possibilities to reuse components from the desktop version of \gls{AF3} during the construction of the \gls{MVP} \gls{AF3} prototype is discussed, which was of particular interest for this case study.

In general, the metamodel, specific model validations, model transformations, and code generators as well as business logic and metamodel-based analyses are primary candidates for easy reuse in the implementation of the model server of the web-based tool. These elements are characterized as being part of the user-interface-independent parts of the desktop tool implementation. Clearly, the user interface had to be renewed, but using abstraction as provided by the \gls{CTK4} layer, e.g., the property section view and its extendable service, can help by providing similar abstractions that are also present in the desktop tooling kernel infrastructure.

In case of the \gls{MVP} discussed in the paper, only the metamodel and a validation implementation was reused (code generators and other business logic and analyses were out of scope for this case study).

\section{Acknowledgment}
\label{sec:acknowledgment}
This study is based upon work supported by the Germany Ministry for Education and Research (BMBF) in the WEBMODEL project (contract number 01IS19034B).

\vspace{-0,14cm}
\bibliographystyle{unsrtnat}
\bibliography{references}  

\end{document}

%% file: acronyms.tex
\newacronym{MVP}{MVP}{Minimal Viable Product}
\newacronym{EMF}{EMF}{Eclipse Modeling Framework\protect\footnote{\protect\url{https://www.eclipse.org/modeling/emf/}}}
\newacronym{AF3}{AF3}{AutoFOCUS~3\protect\footnote{\protect\url{https://af3.fortiss.org}}}
\newacronym{CRUD}{CRUD}{\textit{create, read, update, delete}}
\newacronym{CTK4}{CTK4}{Cloud Tooling Kernel~4}
\newacronym{GLSP}{GLSP}{Graphical Language Server Platform\protect \footnote{\protect \url{https://www.eclipse.org/glsp/}}}
\newacronym{KISS}{KISS}{Kernel Introspection System Service}
\newacronym{SVG}{SVG}{Scalable Vector Graphics}